\documentclass[12pt]{article}

\usepackage[left=2.50cm, right=2.50cm, top=2.50cm, bottom=2.50cm]{geometry}
\usepackage[utf8]{inputenc}
\usepackage[english]{babel}
\usepackage{physics}
\usepackage{hyperref}
\usepackage{amsthm}
\usepackage{mathtools}
\usepackage{graphicx}
\usepackage{changepage}
\usepackage{verbatim}
\usepackage{empheq}
\usepackage{xcolor}
\hypersetup{hidelinks}

\begin{document}
\noindent

{\bf
{\Large 
A note on Harmonic Gauge(s) in Massive Gravity
}} 

\vspace{.5cm}
\hrule

\vspace{1cm}

\noindent

{\large\bf{Giulio Gambuti\footnote{\tt giulio.gambuti@ge.infn.it } and Nicola Maggiore\footnote{\tt nicola.maggiore@ge.infn.it }\\[1cm]}}

\setcounter{footnote}{0}

\noindent
{{}Dipartimento di Fisica, Universit\`a di Genova,\\
via Dodecaneso 33, I-16146, Genova, Italy\\
and\\
{} I.N.F.N. - Sezione di Genova\\
\vspace{1cm}

\noindent
{\tt Abstract~:}

\vspace{.5cm}

We consider the harmonic gauge condition in linearized gravity, seen as a gauge theory for a symmetric tensor field. Once the harmonic gauge condition is implemented, as customary, according to the Faddeev-Popov procedure, the gauge fixed action still depends on one gauge parameter. Consequently, the harmonic gauge appears to be a {\it class of conditions}, rather than a particular one. This allows to give a physical motivation for the covariant harmonic gauge(s), which emerges when the gravitational perturbation is given a mass term. In fact, for a particular choice of harmonic gauge, we find a theory of linearized massive gravity displaying five degrees of freedom, as it should, and which is not affected by the vDVZ discontinuity, differently from what happens in the standard Fierz-Pauli theory.

\vspace{\fill}
{\tt Keywords:} Gauge Field Theories, Faddeev-Popov Procedure, Linearized Massive Gravity, vDVZ Discontinuity.
\newpage
\section{Introduction}

In this paper we consider some properties of the harmonic gauge  fixing condition in the context of Linearized Massive Gravity (LMG) theory \cite{Rubakov:2008nh,Hinterbichler:2011tt,deRham:2014zqa}. The harmonic gauge is adopted in Linearized Gravity (LG) indipendently whether the theory is massive or not, but LMG in the harmonic gauge displays some interesting physical properties which we would like to focus on in this Letter. Of course, in gauge field theory the observables, or, more in general, any physical claim, should not depend on the gauge choice, but in certain gauges some physical properties might be more apparent than in others. We believe that this is the case for the harmonic gauge in LMG.  
The theory of LG is obtained as a perturbation of General Relativity (GR) around an arbitrary background metric $g^{(0)}_{\mu \nu}$ \cite{Carroll:2004st}.  For the scope of this paper we will consider a Minkowskian background, i.e. $g^{(0)}_{\mu \nu} =  \eta_{\mu \nu} =$ diag $(-1,1,1,1)$, so that the whole metric can be written as 
\begin{equation}
g_{\mu \nu}(x) = \eta_{\mu \nu} + h_{\mu \nu}(x)\; ,
\end{equation}
and, expanding the Einstein-Hilbert action 
\begin{equation}
S_{EH}[g] = \int \mathrm{d}^4 x\; \sqrt{-g}\;  R
\end{equation}
to second order in the perturbation $h_{\mu \nu}(x)$, one gets the action 
\begin{equation}\label{intro:inv}
S_{inv}[h]= \int \mathrm{d^4 x} \; \;  \Big[ \; \frac{1}{2} h \partial^2 h \;  - \; h_{\mu \nu } \partial^\mu \partial^\nu h \; - \;\frac{1}{2} h^{\mu \nu} \partial^2 h_{\mu \nu}\;+ \;h^{\mu \nu} \partial_\nu \partial^\rho h_{\mu \rho} \; \Big] \; ,
\end{equation}
where $h(x)\equiv\eta^{\mu\nu}h_{\mu\nu}(x)$.
It is well known that \eqref{intro:inv} is the most general action describing a rank-2 symmetric tensor in a Minkowski space-time and invariant under the infinitesimal gauge transformation 
\begin{equation}\label{intro:symmetry}
h_{\mu \nu} \rightarrow h'_{\mu \nu}= h_{\mu \nu} + \partial_\mu \theta_\nu + \partial_\nu \theta_\mu \; ,
\end{equation}
where $\theta(x)$ is a local infinitesimal gauge parameter. 
As pointed out in \cite{Blasi:2015lrg,Blasi:2017pkk}, when building an action for a MG theory, it is necessary first to gauge fix the invariant massless action \eqref{intro:inv}, and, after that, a mass term might be added. We now briefly recall why this is the case.
We remark that even an intrinsically classic theory as linearized gravity needs a well defined generating functional of the Green functions, without which, for instance, the propagator does not exist, nor, consequently, the corresponding dynamical theory. Adding a mass term directly to the invariant action \eqref{intro:inv}, as done, for instance, in the Fierz-Pauli (FP) approach to MG \cite{Fierz:1939ix}, has a few fundamental flaws.
Firstly, the mass term plays the primary and inappropriate role of gauge fixing the action, breaking the symmetry \eqref{intro:symmetry}. In fact, the FP mass term allows to define a propagator, but this trades the mass for a gauge fixing parameter, which is not physical, in contrast with the fact that mass should be an observable quantity.
Secondly, the FP theory does not display a good massless limit since, at vanishing masses,
one is left with the invariant action $S_{inv}[h]$ \eqref{intro:inv}, which has no propagator for the symmetric tensor field $h_{\mu \nu}(x)$. 
Furthermore, the gravitational couplings predicted by the FP action in the massless limit are inconsistent with those of GR. This fact is known as the vDVZ discontinuity \cite{vanDam:1970vg,Zakharov:1970cc}. A striking effect of the discontinuity is that, in the limit of small mass of the FP theory, the predicted gravitational bending of light due to the presence of a massive body, $e.g.$ a star, differs by 25 \% from that computed in GR.
If, on the other hand, one proceeds as it is customary for a gauge field theory, $i.e.$ by fixing the  gauge before adding a mass term to the action, when taking the zero-mass limit one is left with a well defined theory, for which a propagator exists. 
Therefore, in order to have a MG theory with physical mass parameters and a good massless limit, it is necessary, as a preliminary step, to choose a gauge.
In the context of LG many gauge fixing conditions can be used \cite{Carroll:2004st}. Popular choices are:
\begin{itemize}
\item Transverse gauge 
\item Synchronous gauge
\item Harmonic gauge.
\end{itemize}		
Amongst these, the harmonic gauge condition is the only Lorentz invariant and, written in terms of $h_{\mu \nu}(x)$ and its trace $h(x)$, it reads 
\begin{equation}\label{intro:harmonic}
\partial^\mu h_{\mu \nu} - \frac{1}{2}\partial_\nu h = 0 \; .
\end{equation}
In \cite{Blasi:2015lrg, Blasi:2017pkk} a generalization of the harmonic gauge is used, namely
\begin{equation}\label{intro:gauge_kappa}
\partial^\mu h_{\mu \nu} + \kappa \partial_\nu h = 0 \; .
\end{equation}
The aim of this Letter is to point out that the choice of the covariant harmonic gauge \eqref{intro:harmonic} is particularly clever in a theory of LMG since a physical property (the absence of the vDVZ discontinuity) turns out to be apparent in this gauge, while it is hidden otherwise.  It would be interesting to investigate the absence of the vDVZ discontinuity in other gauges, or, even better, to show that this property does not depend on the gauge choice. This, however, goes beyond the scope of this paper, which is focused on the properties of the  harmonic gauge in the framework of LMG.
An important issue which must be faced concerns the number of degrees of freedom (DOFs), which must be five, for a theory describing a massive spin-2 particle. Therefore, we are dealing with a tough task: that of finding a theory of LMG with a well defined propagator, with a regular massless limit, no vDVZ discontinuity and five DOFs. Our claim is that, at least in the harmonic gauge, this is possible.\\

This paper is organized as follows. In Section 2 we realize the harmonic gauge \eqref{intro:harmonic} according to the standard Faddeev-Popov ($\Phi\Pi$) procedure used in gauge field theory, and we discuss the reason why, amongst the class of covariant gauge conditions \eqref{intro:gauge_kappa}, it is peculiar in the context of LG. We will then use it to build a LMG action and subsequently find the propagator of the theory. In Section 3 we use the equations of motion (EOMs) derived from our LMG action to study for which values of the massive parameters it is possible to recover the five DOFs propagated by a spin-2 massive particle. In Section 4 we show that the results obtained in Section 2 and 3 imply the absence of the vDVZ discontinuity in our theory. Our results are summarized and discussed in the concluding Section 5.

\section{Harmonic gauge} 

The customary $\Phi\Pi$ procedure \cite{Faddeev:1967fc} to introduce a gauge fixing condition into a gauge field theory, is to add a gauge fixing term to the action, which, for the class of covariant gauges \eqref{intro:gauge_kappa}, is
\begin{equation}
S_{gf}[h;k,\kappa] = - \frac{1}{2k} \int \mathrm{d^4 x} \; \;  \left[ \partial^\mu h_{\mu \nu} + \kappa \partial_\nu h \right]^2 \; ,
\label{gfterm}\end{equation}
where $k$ and $\kappa$ are gauge fixing parameters. The gauge fixed action then reads
\begin{equation}\label{1:FP}
S[h;k,\kappa] = S_{inv}[h] + S_{gf}[h;k,\kappa] \; ,
\end{equation}
the ghost sector being factorized out since LG is an abelian gauge theory, and therefore the ghosts are decoupled from the gauge field $h_{\mu\nu}(x)$, as it happens in the Maxwell theory of electromagnetism.
As noticed, in LG the covariant gauge \eqref{intro:gauge_kappa} is realized by means of $two$ gauge parameters: $k$ and $\kappa$. The harmonic gauge \eqref{intro:harmonic}, which is obtained from \eqref{intro:gauge_kappa} by chosing $\kappa=-\frac{1}{2}$, should therefore be thought of as a class of choices, rather than a particular one, corresponding to generic $k$. We shall come again on this point later. In this Letter we are interested in the particular $k= \kappa=-\frac{1}{2}$ harmonic gauge, to which corresponds the gauge fixed action 
 \begin{equation}\label{1:direct_action}
S[h;k= \kappa=-\frac{1}{2}]= \int \mathrm{d^4 x} \; \;  \Big[ \; \frac{1}{4} h \partial^2 h \; - \;\frac{1}{2} h^{\mu \nu} \partial^2 h_{\mu \nu}\; \Big] \; .
\end{equation}
 
The most general mass term which can be added to the action $S$  is 
\begin{equation}\label{1:mass}
S_m[h;m_1^2,m_2^2] = \int \mathrm{d}^4 x  \left[\frac{1}{2} m_1^2 h_{\mu \nu} h^{\mu \nu} + \frac{1}{2} m_2^2 h^2 \right]\; ,
\end{equation}
where $m_1^2$ and $m_2^2$ are massive parameters. The whole LMG action is therefore given by 
\begin{equation}\label{1:MG_action}
S_{MG}[h;m_1^2,m_2^2] = S[h;k= \kappa=-\frac{1}{2}] + S_m[h;m_1^2,m_2^2] \; .
\end{equation}
The action \eqref{1:MG_action} in momentum space reads
\begin{equation} \label{quadr_action_mass}
    S_{MG}[\Tilde{h};m_1^2,m_2^2]= \int \mathrm{d^4 p} \quad \Tilde{h}_{\mu \nu} (-p) \: \Omega^{\mu \nu, \alpha \beta}(p;m_1^2,m_2^2) \:\Tilde{h}_{\alpha \beta}(p)\; ,
\end{equation}
where $\tilde{h}_{\mu \nu}(p)$ is the Fourier transform of $h_{\mu \nu}(x)$, and the kinetic operator $\Omega$ is 
\begin{equation}
   \Omega_{\mu \nu,\alpha \beta}(p;m_1^2,m_2^2) = \frac{1}{2} \left( m_2^2 -  \frac{1}{2} p^2 \right) \eta_{\mu \nu}\eta_{\alpha \beta} +   \frac{1}{2}(p^2 + m_1^2)\mathcal{I}_{\mu \nu,\alpha \beta}  \; ,
\end{equation}
where $\mathcal{I}$ is the rank-4 tensor identity
\begin{equation}\label{1:identity}
            \mathcal{I}_{\mu \nu, \rho \sigma} = \frac{1}{2} (\eta_{\mu \rho} \eta_{\nu \sigma} + \eta_{\mu \sigma} \eta_{\nu \rho}) \; .
\end{equation}
The propagator $G_{\alpha \beta, \rho \sigma}(p;m_1^2,m_2^2)$ is defined by the following equation
\begin{equation}
    {\Omega_{\mu \nu}}^{\alpha \beta}G_{\alpha \beta, \rho \sigma} = \mathcal{I}_{\mu \nu, \rho \sigma}
\end{equation}
which gives
\begin{equation}\label{1:propagator}
    \left<\tilde{h}_{\mu \nu} \tilde{h}_{\alpha \beta} \right>(p) =  G_{\mu \nu, \alpha \beta}(p;m_1^2,m_2^2) = \frac{2}{p^2 +m_1^2} \left[ \mathcal{I}_{\mu \nu, \alpha \beta} - \frac{1}{2} \frac{p^2 - 2m_2^2}{p^2 - m_1^2 - 4m_2^2} \eta_{\mu \nu} \eta_{\alpha \beta}\right] \; .
\end{equation}
Note that the propagator \eqref{1:propagator} displays a good massless limit ($m_1, m_2 \rightarrow 0 $), as expected.\\
\paragraph{Remark} {\small It is worth to point out that plugging the harmonic gauge condition \eqref{intro:harmonic}  directly into the invariant action \eqref{intro:inv},
one obtains the gauge fixed action \eqref{1:direct_action}, $i.e.$ the one coming from the $\Phi\Pi$ gauge fixing term \eqref{gfterm} with $k=- \frac{1}{2}$ and $\kappa=- \frac{1}{2}$. We give now an intuitive explanation of this fact.
Defining the \textit{trace-reversed} field
\begin{equation}\label{2:trace_inv}
\bar{h}_{\mu \nu}\equiv h_{\mu \nu} - \frac{1}{2}\eta_{\mu \nu} h \; ,
\end{equation}
the invariant action \eqref{intro:inv} can be written as 
\begin{equation}\label{1:harm_action}
S_{inv}[\bar{h}] = \int \mathrm{d^4 x} \; \;  \Big[ \; \frac{1}{4} \bar{h} \partial^2 \bar{h} \;  - \;\frac{1}{2} \bar{h}^{\mu \nu} \partial^2 \bar{h}_{\mu \nu}\;+ \;\bar{h}^{\mu \nu} \partial_\nu \partial^\rho \bar{h}_{\mu \rho} \; \Big] \; ,
\end{equation}
and the harmonic gauge condition \eqref{intro:harmonic} reduces to 
\begin{equation}\label{1:harm_bar}
\partial^\mu \bar{h}_{\mu \nu}= 0\; . 
\end{equation}

In terms of $\bar{h}_{\mu \nu}(x)$, the gauge symmetry \eqref{intro:symmetry} writes 

\begin{equation}
\bar{h}_{\mu \nu} \rightarrow \bar{h}'_{\mu \nu}= \bar{h}_{\mu \nu} + \partial_\mu \theta_\nu + \partial_\nu \theta_\mu  - \eta_{\mu \nu} \partial_\alpha \theta^\alpha \; .
\end{equation}

In fact, $S_{inv}[\bar{h}] = S_{inv}[\bar{h}']$ and, in close analogy with Maxwell theory, as long as $\partial^2 \theta_\mu(x) = 0$, the transformed field $\bar{h}'_{\mu \nu}(x)$ satisfies equation \eqref{1:harm_bar}. 
Looking at the action \eqref{1:harm_action}, we see that only the third term involves $\partial^\mu \bar{h}_{\mu \nu}(x)$. So it is clear that, in the EOMs, the terms proportional to $\partial^\mu \bar{h}_{\mu \nu}(x)$ come from the variation of 

\begin{equation}
\int \mathrm{d}^4x \; \; \bar{h}^{\mu \nu} \partial_\nu \partial^\rho \bar{h}_{\mu \rho} \; . 
\end{equation}

Plugging \eqref{1:harm_bar} into the action \eqref{1:harm_action} results in eliminating this term and therefore every term in the EOMs containing $\partial^\mu \bar{h}_{\mu \nu}(x)$ vanishes. The key observation is that we would have obtained the same result by substituting the constraint \eqref{1:harm_bar} into the EOMs obtained from the invariant action \eqref{1:harm_action}.
Therefore, plugging the harmonic gauge condition \eqref{intro:harmonic} directly into the action is a valid way of gauge fixing, since this peculiarly corresponds to a particular choice $(k=\kappa=-\frac{1}{2})$ of the $\Phi\Pi$ gauge fixing term \eqref{gfterm}.

}

\section{EOMs and DOFs in the harmonic gauge}  

In their original paper [7] on the theory of a massive spin-$f$ particle,  the issue faced by Fierz and Pauli in the case $f=2$ was to find a theory characterized by the following constraints: 
\begin{equation}
(\partial^2 - m_1^2) h_{\mu \nu} = 0\ ,
\label{KKeq}\end{equation}
and
\begin{eqnarray}
 \partial^\mu h_{\mu \nu}&=&0 \label{transv}\\
 h &=& 0\label{traceless}\ .
 \end{eqnarray}
Eq. \eqref{KKeq} is the massive wave equation for a massive rank-2 symmetric tensor field, and Eqs. \eqref{transv} and \eqref{traceless} represent the five constraints (respectively transversality and tracelessness) which lower the number of independent components of $h_{\mu\nu}(x)$ from ten to five, as it should be for a massive spin-2 particle. In this way Fierz and Pauli recovered five DOFs from a theory which otherwise concerns ``only'' the ten components of a generic symmetric 4D rank-2 tensor field. In this Section we show that, at least in the case of the harmonic gauge choice, five DOFs are indeed obtained, which is the necessary preliminary condition which must be fulfilled before dealing with any further question. We follow here the same steps as in the original paper \cite{Fierz:1939ix}. 
The action $S_{MG}$ \eqref{1:MG_action} implies the following EOM
\begin{equation}\label{2:eom}
(- \partial^2 + m_1^2) h_{\mu \nu} + \left( \frac{1}{2}\partial^2 + m_2^2\right) \eta_{\mu \nu} h=0 \; .
\end{equation}
Taking the trace of \eqref{2:eom} we get
\begin{equation}
(\partial^2 + m_1^2 + 4m_2^2) h= 0 \; ,
\label{2:trace}\end{equation}
while saturating \eqref{2:eom} with $\partial^\nu$ and using the harmonic gauge fixing condition \eqref{intro:harmonic} we find
\begin{equation}
\left( \frac{m_1^2}{2}  + m_2^2 \right) \partial_\mu h = 0 \; . 
\label{2:derivative}
\end{equation}
Looking at \eqref{2:derivative}, if $m_2^2 \neq - \frac{m_1^2}{2}$  we obtain $\partial_\mu h(x) = 0$, which implies the transversality condition \eqref{transv} because of the harmonic gauge condition \eqref{intro:harmonic}, and inserting it into \eqref{2:trace} we get 
\begin{equation}
(m_1^2 + 4m_2^2) h= 0\; . \label{2:m1+4m2} 
\end{equation}
If $ m_1^2 + 4m_2^2 \neq 0 $, then \eqref{2:m1+4m2} implies the tracelessness condition $h(x)=0$ \eqref{traceless}, which, plugged into \eqref{2:eom} yields the massive wave equation \eqref{KKeq}.
 If instead $m_1^2 + 4m_2^2 = 0$, Eq. \eqref{2:eom} and $\partial_\nu h(x) = 0$ imply 
\begin{equation}
(\partial^2 - m_1^2) H_{\mu \nu} = 0 \; ,
\end{equation}
where $H_{\mu \nu}(x)$ is the traceless part of $h_{\mu \nu}(x)$
\begin{equation}
H_{\mu \nu}\equiv   h_{\mu \nu} - \frac{1}{4}\eta_{\mu \nu} h \; ,
\end{equation}
which satisfies $\partial^\mu H_{\mu \nu}(x)=0$. 
Finally, considering the case $m_2^2 = - \frac{m_1^2}{2}$, Eq. \eqref{2:trace} becomes 
\begin{equation}
(\partial^2 - m_1^2) h = 0 \; ,
\end{equation}
which inserted into the EOMs \eqref{2:eom} gives again the wave equation \eqref{KKeq}.\\

The results of this section are summarized in the following table:

\begin{center}
\begingroup
\setlength{\tabcolsep}{10pt} 
\renewcommand{\arraystretch}{1.5} 
\begin{tabular}{ c||c|c|c } 
 Mass Condition & Propagation & Constraints & DOFs \\[6pt]
 \hline\hline
 $m_2^2 = - \frac{m_1^2}{2}$    & $(\partial^2 - m_1^2) h_{\mu \nu} = 0$     & $\partial^\mu h_{\mu \nu} =0$      &  6 \\[6pt]
 \hline
 $m_2^2 =  -\frac{m_1^2}{4}$      & $(\partial^2 - m_1^2) H_{\mu \nu} = 0  $     &  $\partial^\mu H_{\mu \nu}=0$      &  6 \\ [6pt]
  \hline
 $m_2^2 \neq - \frac{m_1^2}{2} \: \& \: m_2^2 \neq  -\frac{m_1^2}{4}$    & $(\partial^2 - m_1^2) h_{\mu \nu} = 0$ & $\partial^\mu h_{\mu \nu}=0$, $h=0$      &    5 \\[6pt] 
\end{tabular}
\endgroup

\vspace{.2cm}
{\footnotesize {\bf Table 1.}
Mass conditions and corresponding DOFs.}

\end{center}

Only the third mass condition listed in Table 1 guarantees that the propagating graviton $h_{\mu\nu}(x)$ displays 5 DOFs, as expected for a massive spin-2 field. Notice that the FP prescription $m_1^2+m_2^2=0$ does satisfy this requirement: the FP theory, indeed, has been introduced in order to get the correct number of DOFs for MG \cite{deRham:2014zqa}. A theory of LMG satisfying the third mass condition, therefore, contains the FP tuning.\\

A comment is in order concerning the comparison with the FP theory. It has always been thought that the FP model is the unique local theory of MG that consistently (apart from the bad massless limit and the vDVZ discontinuity, but without ghost or tachyon instabilities) describes the five DOFs of a massive spin-2 representation of the Poincar\'e group. In particular, modifying the kinetic (derivative) part of MG away from that of general relativity results in ghost instabilities in the helicity-2 sector, which becomes obvious in the high-energy ($i.e.$ massless) limit, where this sector is supposed to reduce to GR, which is basically the statement that GR is the unique inconsistency-free theory of a massless helicity-2 particle. Moreover, adding a mass term other than the one of Fierz and Pauli $m_1^2 + m_2^2 \neq 0$ results in a scalar ghost (the ``sixth'' DOF) \cite{Boulware:1973my}. So the question arises naturally: how comes that this does not happen in the case discussed in this paper, where we just proved that it is possible to go outside the FP tuning, nonetheless keeping five DOFs ? The answer to this legitimate doubt resides in the definition of the physical sector of a gauge field theory.
It is true that we are modifying the kinetic part of MG, but through a gauge fixing term. A true one, not like the FP ``mass'' term. We have in mind the case of any gauge field theory, where the gauge fixing term is necessary to define the partition function (and the Green functions it generates, starting from the propagator). Certainly the kinetic (derivative) part of the theory is modified, but in the non-physical sector (the gauge fixing one). This means that the observables are insensitive to this particular modification. As we are going to show in the next Section, the absence of the vDVZ discontinuity is exactly a consequence of this, since it concerns the continuity with GR of a particular observable (the scattering amplitude of two bodies). Therefore, as in Maxwell or Yang-Mills theory, the gauge fixing term does not change the physics of LG, despite the fact that it modifies its dynamic quadratic part.

\section{Absence of the vDVZ discontinuity} 

The vDVZ discontinuity \cite{vanDam:1970vg,Zakharov:1970cc} is a well known issue affecting the FP theory of MG \cite{Rubakov:2008nh,Hinterbichler:2011tt},
and it is essentially a statement about the number of DOFs of a massive spin-2 particle. In this sense this Section is tightly related to the previous one. Unlike a massless graviton which has two DOFs, a massive graviton has five, as required by the representation theory of the Poincar\'e group. These five DOFs are comprised of the helicity-2 (two DOFs), helicity-1 (two DOFs), and helicity-0 (scalar, one DOF) components. In the short distance/high-energy (``massless'') limit of the theory, where helicity becomes a good quantum number, the helicity-2 component encodes the ``general relativistic'' part of the theory, while the helicity-0 component does not decouple from external sources (as far as the full graviton field couples minimally to matter), leading to a physical ``fifth force''. Hence the ``discontinuity'' of the linear theory.\footnote{We thank the Reviewer for this contribution}
The common way to fix this problem of the FP theory is to adopt the Stueckelberg mechanism \cite{Hinterbichler:2011tt,Stueckelberg:1900zz}, at the rather expensive price of introducing additional fields in order to restore the diffeomorphism invariance broken by the mass terms. In this Section we show that, 
by choosing the harmonic gauge \eqref{intro:harmonic} belonging to the $k=-\frac{1}{2}$ class, the continuity with GR can be restored in a more natural way, without invoking extra fields. 
We will now follow the same steps described in \cite{Rubakov:2008nh}, which lead to the evidence of the presence of the vDVZ discontinuity in the FP theory, to verify, instead, its absence in the LMG with $k=\kappa=-\frac{1}{2}$ harmonic gauge fixing.
As discussed in \cite{Rubakov:2008nh}, the propagator of the FP theory is 
\begin{equation}\label{fp_prop}
    \mbox{FP}: \quad G_{\mu \nu, \alpha \beta} =  \frac{2}{p^2 + m_G^2} \left[\mathcal{I}_{\mu \nu, \alpha \beta} -\frac{1}{3}\eta_{\mu \nu}\eta_{\alpha \beta} + (p\text{-dependent terms}) \right]\; ,
\end{equation}

where $m_G^2 = m_1^2 = -m_2^2$, while in the linearized limit of GR it is 

\begin{equation}\label{massless_prop}
    \mbox{GR}: \quad G_{\mu \nu, \alpha \beta} =  \frac{2}{p^2} \Big[\mathcal{I}_{\mu \nu, \alpha \beta} -\frac{1}{2}\eta_{\mu \nu}\eta_{\alpha \beta} + (p\text{-dependent terms}) \Big]\; .
\end{equation}
In the $k=\kappa=-\frac{1}{2}$ harmonic gauge theory (denoted here by H), the propagator is given by \eqref{1:propagator}:
\begin{equation}
\mbox{H}: \quad G_{\mu \nu, \alpha \beta} = \frac{2}{p^2 +m_1^2} \left[ \mathcal{I}_{\mu \nu, \alpha \beta} - \frac{1}{2} \frac{p^2 - 2m_2^2}{p^2 - m_1^2 - 4m_2^2} \eta_{\mu \nu} \eta_{\alpha \beta}\right] \; .
\end{equation}
Following \cite{Rubakov:2008nh}, the interaction between two non relativistic bodies in the three cases (FP, GR and H) is computed by contracting the propagator with the 00-components of the energy-momentum tensors of the two bodies, $T_{\mu \nu}(x)$ and $T'_{\alpha \beta}(x)$, which are conserved ($i.e.$ $p_\nu \Tilde{T}^{\mu \nu} = 0$). In the non-relativistic case, all other components are negligible. So, taking into account the coupling constants of the three theories, which \textit{a priori} do not coincide, in the massless limit we get the interaction strengths

\begin{align}
    \mbox{FP}: \quad G_{FP}\: \Tilde{T}^{\mu \nu} G_{\mu \nu, \alpha \beta} \Tilde{T}'^{\alpha \beta}=& \frac{4}{3}G_{FP}\: \Tilde{T}_{00}\Tilde{T}'_{00}\frac{1}{p^2} \label{wrong_G} \; ,\\[10pt]
    \mbox{GR}: \quad G_{GR}\: \Tilde{T}^{\mu \nu} G_{\mu \nu, \alpha \beta} \Tilde{T}'^{\alpha \beta}=& G_{GR}\: \Tilde{T}_{00}\Tilde{T}'_{00}\frac{1}{p^2} \; , \\[10pt]
    \mbox{H}: \quad G_{H}\: \Tilde{T}^{\mu \nu} G_{\mu \nu, \alpha \beta} \Tilde{T}'^{\alpha \beta}=& G_{H}\: \Tilde{T}_{00}\Tilde{T}'_{00}\frac{1}{p^2} \; .
\end{align}

The constant $G_{GR}$ is the one that has been experimentally measured to a certain value $G_{GR}\equiv G_{\text{Newton}}$. Imposing that the couplings of FP and H coincide with that of GR, we get

\begin{equation}\label{3:constants}
    \frac{4}{3} G_{FP}= G_{H} = G_{GR} \equiv G_{\text{Newton}}\; .
\end{equation}

We now consider the energy-momentum tensor of an electromagnetic wave (say $T'(x)$), which is traceless, and consider its gravitational interaction with that of a massive body ($T(x)$). In the massless limit we obtain the interaction strengths

\begin{align} 
 \mbox{FP}: \quad G_{FP}\: \Tilde{T}^{\mu \nu} G_{\mu \nu, \alpha \beta} \Tilde{T}'^{\alpha \beta}&= G_{FP}\: \Tilde{T}_{00}\Tilde{T}'_{00}\frac{2}{p^2}=  \frac{4}{3}G_{Newton}\: \Tilde{T}_{00}\Tilde{T}'_{00}\frac{2}{p^2}  \label{FP} \; , \\[10pt]
    \mbox{GR}: \quad G_{GR}\: \Tilde{T}^{\mu \nu} G_{\mu \nu, \alpha \beta} \Tilde{T}'^{\alpha \beta}&= G_{GR}\: \Tilde{T}_{00}\Tilde{T}'_{00}\frac{2}{p^2}= G_{Newton}\: \Tilde{T}_{00}\Tilde{T}'_{00}\frac{2}{p^2}  \label{GR} \; , \\[10pt]
   \mbox{H}: \quad G_{H}\: \Tilde{T}^{\mu \nu} G_{\mu \nu, \alpha \beta} \Tilde{T}'^{\alpha \beta}&= G_{H}\: \Tilde{T}_{00}\Tilde{T}'_{00}\frac{2}{p^2}=G_{Newton}\: \Tilde{T}_{00}\Tilde{T}'_{00}\frac{2}{p^2}\; ,
\end{align}
where we used Eq. \eqref{3:constants}. It is clear from the above equations that, while the FP interaction differs from that of GR even in the massless limit, the $k=\kappa=-\frac{1}{2}$ harmonic MG result happily matches that of GR. 

\section{Summary and Discussion}

In this Letter we highlighted a few properties of the covariant harmonic gauge which, as far as we know, have not been remarked elsewhere, although, in our opinion, deserve some attention. In textbooks, the harmonic, called also Lorenz, gauge is presented as a covariant condition on the gravitational perturbation	$h_{\mu\nu}(x)$ which just simplifies the linearized Einstein equations (see for instance \cite{Carroll:2004st}). We believe that, besides this rather formal motivation, this choice can be  strengthened by more physical motivations. Our observation comes from the close analogy between LG and ordinary gauge field theory. 
The approach of our paper, not that revolutionary indeed, is to treat the action (3) of Linearized Gravity as an ordinary free gauge field theory, like Maxwell theory, which, hence, needs a gauge fixing in order to be defined. Only after that, a mass term is added, thus avoiding {\it ab initio} the divergent massless limit. In this way the mass term is allowed to be a true one, without playing the double role of mass and gauge fixing term, as it happens in the FP theory.  The FP tuning $m^2_1 + m^2_2 = 0$ is mandatory if the mass term is forced to serve as a gauge fixing too, as it is known. But, if LG is modified in its non-physical sector by a gauge fixing term (and the absence of the vDVZ discontinuity encourages to believe that this is the case), the mass parameters turn out to be less constrained, being granted that the DOFs are five, as it should and as we proved.
As it is well known, the standard way to restrict the space of connections in order to eliminate the redundancy of the Green functions generating functional $Z[J]$, is the $\Phi\Pi$ procedure \cite{Faddeev:1967fc}, which consists in introducing in the path integral a $\delta$-functional, which is then exponentiated in order to be able to deal with a gauge fixed action. In the case of Maxwell theory, we have
\begin{equation}
Z^{(A)}[0] = \int DA_\mu\; e^{iS_{inv}[A]	}    \rightarrow
\int DA_\mu\; \delta(F[A] -f(x))\;  e^{iS_{inv}[A]	} 
\propto
\int DA_\mu\;    e^{iS_{inv}[A]-\frac{i}{2\xi}\int (F[A])^2}\; ,
\label{zmax}\end{equation}
where $F[A]=\partial A(x) $ identifies the covariant Lorenz gauge fixing condition, which is the unique linear possibility for the gauge field $A_\mu(x)$, and $\xi$ is the gauge parameter, which is unique as well. Taking, as it is, LG as a gauge field theory for a symmetric tensor $h_{\mu\nu}(x)$, the realization of the harmonic covariant gauge fixing condition \eqref{intro:gauge_kappa} should go analogously as follows
\begin{equation}
Z^{(h)}[0] = \int Dh_{\mu\nu}\; e^{iS_{inv}[h]	}    \rightarrow
\int Dh_{\mu\nu}\; \delta(F[h;\kappa] -f(x))\;  e^{iS_{inv}[h]	} 
\propto
\int DA_\mu\;    e^{iS_{inv}[h]-\frac{i}{2k}\int (F[h;\kappa])^2}\; ,
\label{zh}\end{equation}
where $S_{inv}[h]$ is the LG action \eqref{intro:inv}, and $F[h;\kappa]$ is the (vectorial) covariant gauge fixing condition on $h_{\mu\nu}(x)$ \eqref{intro:gauge_kappa}. The $\Phi\Pi$ procedure allows to have a well defined theory, with a propagator for the field $h_{\mu\nu}(x)$, which eventually can be given a mass, as discussed previously. As we remarked,  the gauge fixing term \eqref{gfterm} depends on $two$ gauge parameters, $k$ and $\kappa$. The gauge parameter $k$ in \eqref{zh} (and in \eqref{gfterm}) is the analogous of $\xi$ for the Maxwell theory \eqref{zmax}: it defines, for instance, the Landau ($k=0$) gauge. We therefore might consider $k$ as a kind of ``primary'' gauge parameter, which is then fine-tuned by $\kappa$, which appears to be a ``secondary'' gauge parameter. We remark that, by substituting directly the harmonic gauge fixing condition \eqref{intro:gauge_kappa} into the action $S_{inv}[h]$ \eqref{intro:inv}, this non trivial structure of the gauge fixing is lost.  According to this approach (rather standard, actually, despite the fact that it seems to be uncommon in the context of LG), one should refer to the harmonic gauge 
$\kappa=-\frac{1}{2}$ as a {\it class of choices}, rather than to a particular one. For instance, it makes sense to speak of a harmonic-Landau gauge, which corresponds to $k=0$ and $\kappa=-\frac{1}{2}$, or harmonic-Feynman gauge ($k=1,\ \kappa=-\frac{1}{2}$), and so on. In this Letter we hope to have given some insights concerning the meaning of the harmonic gauge in LG, motivating this choice in the case of a particular harmonic gauge, the one belonging to the larger $k=-\frac{1}{2}$ class. In this gauge, in fact, the five DOFs of the graviton can be given physical masses, which do not serve, as in the FP theory of LMG, as gauge parameters. 
The theory of LMG corresponding to the third condition in Table 1, which contains as a particular case the FP mass term,
displays, we might say by construction, a good massless limit and, remarkably, it is not affected by the vDVZ discontinuity. Of course this physical property should depend on the particular mass term which has been chosen, and not on the gauge choice. Our claim is that, as it often happens in gauge field theory, in the particular $k=\kappa=-\frac{1}{2}$ - harmonic gauge, this property is apparent, differently from what happens in the standard FP theory, where it is absent, as it is well known. Probably the deep reason for this flaw is the fact the FP mass serves, quite unnaturally, as gauge fixing tool to allow the existence of a propagator, which nonetheless lacks a good massless limit, more than a real mass parameter.

\end{document}